\documentclass[12pt,preprint]{aastex}

\def\ie{{\it i.e.} }                    
\def\a4{\hsize 17.0cm \vsize 25.cm}

\shorttitle{2D Model of a Bimodal Wind}
\shortauthors{ W\"unsch et al.}

\begin{document}

\title{Two-Dimensional Hydrodynamic Models of Super Star Clusters
with a Positive  Star Formation Feedback}

\author{R. W\"unsch \altaffilmark{1}}
\affil{Cardiff University, Queens Buildings, The Parade, Cardiff. CF24 3AA,
United Kingdom; richard@wunsch.cz}
\altaffiltext{1}{Astronomical Institute, Academy of Sciences of the Czech
Republic, Bo\v{c}n\'\i\ II 1401, 141 31 Prague, Czech Republic}
\author{G. Tenorio-Tagle}
\affil{Instituto Nacional de Astrof\'\i sica Optica y
Electr\'onica, AP 51, 72000 Puebla, M\'exico}
\author{J. Palou\v{s}}
\affil{Astronomical Institute, Academy of Sciences of the Czech
Republic, Bo\v{c}n\'\i\ II 1401, 141 31 Prague, Czech Republic}
\and
\author{S. Silich}
\affil{Instituto Nacional de Astrof\'\i sica Optica y
Electr\'onica, AP 51, 72000 Puebla, M\'exico}

\begin{abstract}
Using the hydrodynamic code ZEUS, we perform 2D simulations to determine the
fate of the gas ejected by massive stars within super star clusters. It turns
out that the outcome depends mainly on the mass and radius of the cluster. In
the case of less massive clusters, a hot high velocity ($\sim 1000$ km s$^{-1}$)
stationary wind develops and the metals injected by supernovae are dispersed to
large distances from the cluster. On the other hand, the density of the
thermalized ejecta within massive and compact clusters is sufficiently large as
to immediately provoke the onset of thermal instabilities. These deplete,
particularly in the central densest regions, the pressure and the pressure
gradient required to establish a stationary wind, and instead the thermally
unstable parcels of gas are rapidly compressed, by a plethora of re-pressurizing
shocks, into compact high density condensations. Most of these are unable to
leave the cluster volume and thus accumulate to eventually feed further
generations of star formation. 

The simulations cover an important fraction of the  parameter-space, which
allows us to estimate the fraction of the reinserted gas which accumulates
within the cluster and the fraction that leaves the cluster as a function of
the cluster mechanical luminosity, the cluster size and heating efficiency.
\end{abstract}


\section{Introduction}

Due to their stellar mass, which in some cases amounts to several million
M$_\odot$, and their compactness, as they only span a few parsecs, young ($<
10$~Myr) super star clusters (SSCs) represent the most spectacular and dominant
mode of star formation in  starburst and interacting galaxies
\citep{1995ApJ...446L...1O,1997RMxAC...6....5H,1995AJ....109..960W,2005ApJ...619..270M}.
SSCs have been detected in the optical, UV and X-rays, and also in the radio
continuum and IR regimes, as some of them are deeply embedded behind dense
obscuring material, leading to powerful ultra-dense HII regions
\citep{1999ApJ...527..154K,2007ApJ...668..168G}.

On theoretical grounds, it has been inferred that such extreme modes of
star formation should  lead to several tens of thousands of massive stars, all
of them known to rapidly ($\leq 50$~Myr) reinsert, through stellar winds and
supernova (SN) explosions,  a large fraction of their mass back into the ISM.
The first (adiabatic) approach to the hydrodynamics of the matter reinserted
within a SSC is due to \citet[hereafter CC85]{1985Natur.317...44C}. In their
model, the stellar sources of mass and energy, are assumed to be equally
spaced within the SSC volume of radius $R_\mathrm{SC}$. They also assumed that
all of the kinetic energy provided by massive stars is immediately, and in situ,
thermalized via random collisions of the ejecta from neighboring sources.
This results in energy and mass deposition rate densities $q_e =
(3L_\mathrm{SC})/(4\pi R_\mathrm{SC}^3)$ and $q_m = (3\dot{M}_\mathrm{SC})/(4\pi
R_\mathrm{SC}^3)$, respectively, where $L_\mathrm{SC}$ and $\dot{M}_\mathrm{SC}$
are the cluster mechanical luminosity and mass deposition rates. These
assumptions lead to a high temperatures gas ($T > 10^7$~K) in which the
interstellar cooling law is close to its minimum value, and this justifies
their adiabatic assumption. In the adiabatic model of CC85, the  thermalized hot
gas  rapidly settles  into almost constant density and  temperature
distributions, although a slight outward pressure gradient establishes a
particular velocity distribution with the stagnation point (\ie zero
velocity) at the cluster center. The gas velocity increases then almost
linearly with radius to reach the sound speed ($c_\mathrm{SC}$) right at the
cluster surface and then streams into the surrounding lower pressure ambient
medium to reach its terminal velocity ($v_\mathrm{A\infty} \sim 2 c_\mathrm{SC}$),
while the density and temperature of the outflow, the cluster wind, decrease as
$r^{-2}$ and $r^{-4/3}$, respectively. The solution of such a stationary outflow
depends on  three variables: the cluster size ($R_\mathrm{SC}$), the mass
deposition rate ($\dot M_\mathrm{SC}$) and the mechanical luminosity of the
cluster ($L_\mathrm{SC}$). Knowledge of these three variables allows one to
solve the hydrodynamic equations analytically and workout the run of density,
temperature and velocity of the stationary outflow. Note that $\dot
M_\mathrm{SC}$ is usually replaced by the adiabatic terminal speed ($v_\mathrm{A\infty}
=  (2 L_\mathrm{SC}/\dot M_\mathrm{SC})^{1/2}$).

The model then yields a stationary flow in which the matter reinserted by the
evolving massive stars ($\dot M_\mathrm{SC}$) equals the amount of matter
feeding the cluster wind ($4 \pi R_\mathrm{SC}^2 \rho_\mathrm{SC}
c_\mathrm{SC}$); where $\rho_\mathrm{SC}$ is the reinserted gas density value at
the star cluster surface. As $L_\mathrm{SC}$ and $\dot M_\mathrm{SC}$ increase
linearly with the cluster mass ($L_\mathrm{SC} \sim M_\mathrm{SC}$,
$\dot{M}_\mathrm{SC} \sim M_\mathrm{SC}$), the adiabatic model predicts that the
more massive a cluster is, the more powerful is its resultant wind. This latter
conclusion breaks down if one relaxes the adiabatic assumption
\citep[see][]{2004ApJ...610..226S}. More massive clusters deposit larger
amounts of matter and thereby deliver a sufficiently high density,
$\rho_\mathrm{SC} = \dot M_\mathrm{SC} / (4 \pi R_\mathrm{SC}^2)$, to provoke
strong radiative cooling. Thus, as radiative cooling ($Q$ = $n^2 \Lambda(T,
Z)$; where $\Lambda(T, Z)$ is the cooling function) is proportional to
$M_\mathrm{SC}^2$ and $L_\mathrm{SC}$ is proportional to $M_\mathrm{SC}$,
there is a threshold mechanical luminosity (for given $R_\mathrm{SC}$ and
$\dot{M}_\mathrm{SC}$), above which strong radiative cooling takes over despite
the large temperatures and the minimum value of the interstellar cooling law at
these temperatures.

Note that details of the thermalization process have been largely ignored
although more recent formulations of the problem \citep{2007A&A...471..579W,
2007ApJ...669..952S} have inferred  that a significant fraction of the deposited
mechanical energy could be radiated away as soon as it is inserted. In such
cases, only  a fraction of the deposited energy remains available to  heat the
thermalized matter. This fraction is called the heating efficiency,
$\eta$. It is assumed by different authors to have values between $0.01$
and $1$ \citep{1998A&A...337..338B,2004A&A...424..817M}, and shown to acquire
small values in the case of massive compact SSCs
\citep[see][]{2007ApJ...669..952S}.

\begin{figure}
\plotone{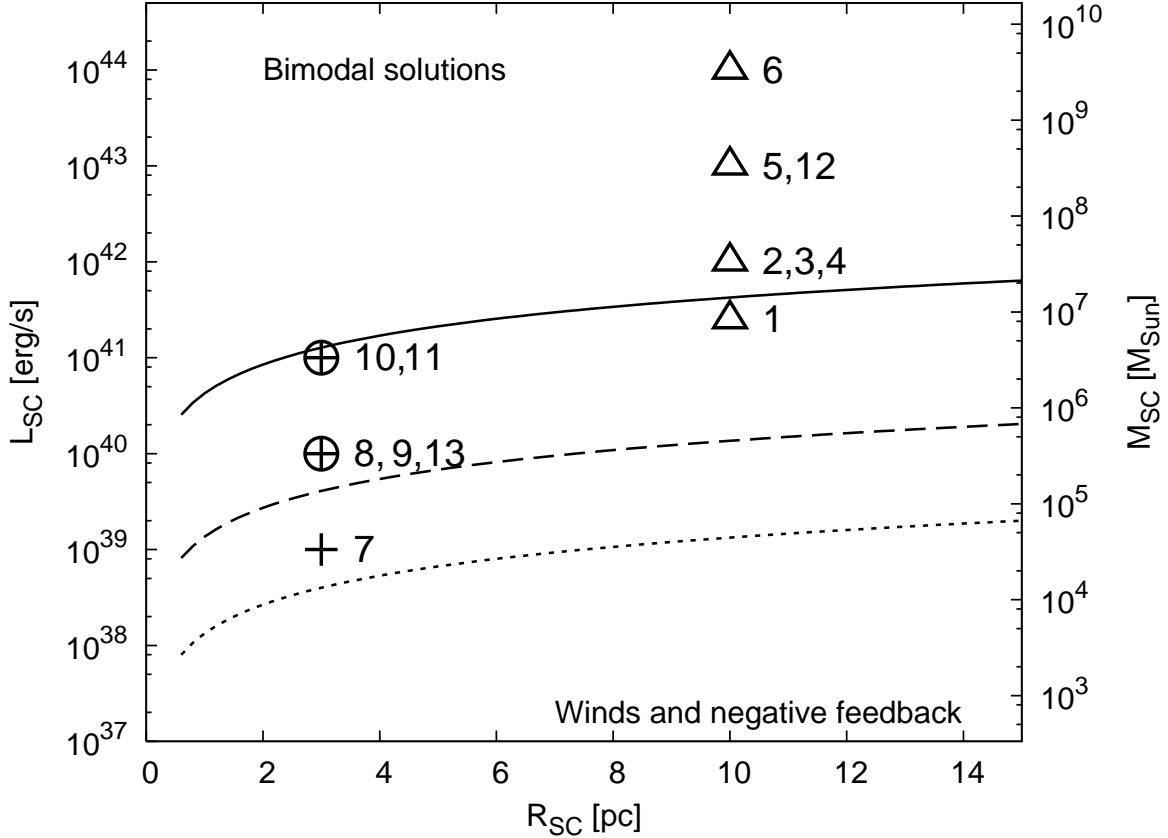}
\caption{The threshold lines. 
Clusters above the threshold line evolve in a bimodal regime in 
which the injected matter is accumulated inside their densest inner 
regions while the outer zones develop a strongly radiative stationary 
wind. The threshold lines were calculated for the adiabatic terminal velocity
$V_\mathrm{A\infty} = 1000$~km s$^{-1}$ and three different heating efficiencies
$\eta = 1.0, 0.3$ and $0.1$ denoted with solid, dashed and dotted lines, 
respectively. The 2D simulations (see Table~\ref{modtab}) are represented by 
symbols. Different symbols denote different heating efficiencies (and thus 
are to be compared with the corresponding threshold  line) $\eta = 1.0$ 
(triangles), $\eta = 0.3$ (circles), and $\eta = 0.1$ (plus signs).
The secondary y-axis shows the approximate mass of the cluster obtained using a
relation $M_\mathrm{SC} = (L_\mathrm{SC} / 3\times 10^{40}) 10^6$~M$_\odot$
\citep{1999ApJS..123....3L}.}
\label{Lcrit}
\end{figure}

\begin{figure}
\plotone{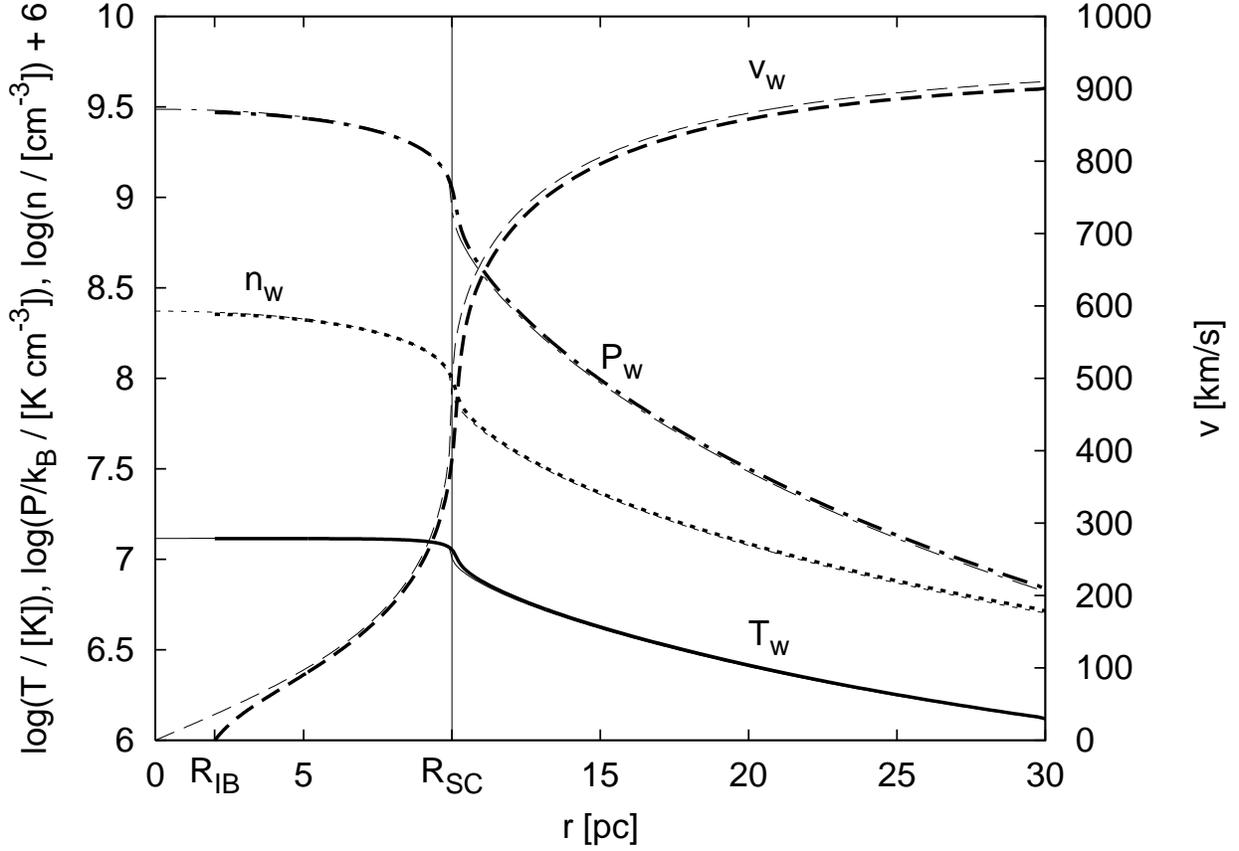}
\caption{Model 1 ($L_\mathrm{SC}/L_\mathrm{crit} = 0.5$): radial profiles of 
the wind particle density, $n_w$ (note the units are $m^{-3}$ to
fit within the figure), temperature ($T_w$), pressure ($P_w$) and radial
velocity ($v_w$). The thin lines represent the semi-analytical solution 
\citep{2004ApJ...610..226S}, the
thick lines are results from the 2D simulation (model 1) at $t = 0.25$~Myr.
The simulation, after a short initial relaxation period, 
stays perfectly stationary and spherically
symmetric at all  times.}
\label{m1d}
\end{figure}

Figure~\ref{Lcrit} presents the threshold mechanical luminosity found by
\citet{2004ApJ...610..226S}, here re-calculated for three different values of
the heating efficiency $\eta$. Clusters with a mechanical luminosity (or mass)
far below this line evolve in the quasi-adiabatic regime.  For these,  the
Chevalier \& Clegg model provides a good approximation to the structure and
hydrodynamics of the star cluster wind
\citep{2000ApJ...536..896C,2001ApJ...559L..33R}. Figure~\ref{m1d} displays
radial profiles of temperature, particle density, pressure and velocity, typical
of such  steady state winds. 

For clusters with a mechanical luminosity  close to the threshold value, the
temperature distribution within the stationary wind becomes very different from
that predicted by the adiabatic model \citep{2004ApJ...610..226S}. This is
because, as soon as the temperature of the wind decreases to approximately $\sim
10^6$~K, radiative  cooling begins to increase sharply (mainly due to f-b and
b-b transitions of  heavy elements) and the temperature in the free wind region
drops rapidly  several orders of magnitude. Nevertheless such clusters manage to
eject all the matter deposited by stellar winds and supernovae, and thereby
sustain a stationary wind.

\begin{figure}
\plotone{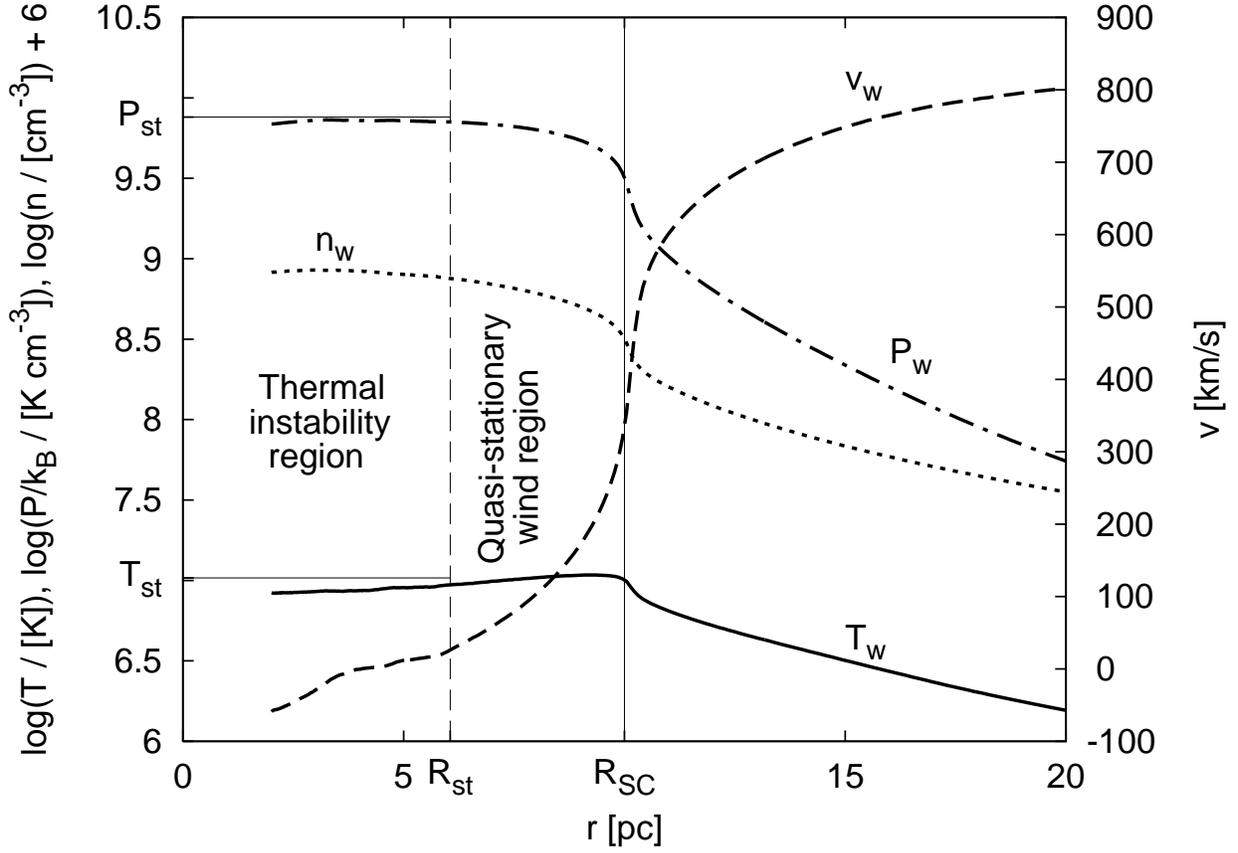}
\caption{The internal structure of SSCs in the bimodal regime.
The cluster volume presents  two distinct regions. Below the
stagnation radius $R_\mathrm{st}$: is the  thermally  unstable region and 
above it lies the outer quasi-stationary wind region.
The radial profiles of the wind particle density, temperature, pressure and
radial velocity were obtained as the axial time averages values of the 2D
simulation (model~3) between $0.4$ and $0.8$~Myr. Only the hot gas 
($T > 2\times 10^4$) was taken into consideration for this plot. The values 
of temperature and pressure at the stagnation radius, obtained from the 
semi-analytical model (thin solid lines across the inner region), are given with
horizontal lines.}
\label{figbm}
\end{figure}

For clusters above the threshold line, radiative cooling becomes a dominant
factor. Radiative cooling affects first the central densest regions causing a
sudden drop in pressure. This promotes the shift of the stagnation point out of
the cluster center. Such clusters adhere to a bimodal solution
\citep{2007ApJ...658.1196T,2007A&A...471..579W} in which the stagnation radius
($R_\mathrm{st}$) defines two different regions within the cluster volume (see
Figure~\ref{figbm}). On one hand, there is an outer
shell in which the deposited energy, although affected by strong radiative
cooling, is still able to drive an outward stationary wind, by making the gas
velocity acquire the sound speed exactly at the cluster surface. On the other
hand, the matter deposited inside the stagnation radius is strongly affected by
radiative cooling and becomes thermally unstable. The instability leads to a rapid
and continuous loss of energy from large and small parcels of gas, thereby
reducing their temperature and pressure. These events lead immediately to the
formation of strong shocks that emanate from the hot, high pressure regions and
are driven into the cold parcels of gas in order to restore their pressure.
These have been termed re-pressurizing shocks, and have been invoked in several
astrophysical circumstances, such as the formation of globular clusters
\citep{1995ApJ...442..618V}, and also in the cooling of supernova
matter within superbubbles, leading to highly metallic droplets falling onto the
galaxy \citep{1996AJ....111.1641T}. In the context of the matter cooling
inside the superstar cluster stagnation radius, the re-pressurizing shocks have been
first modelled by means of 1D numerical hydrodynamics in
\citet{2007ApJ...658.1196T}. The re-pressurizing shocks rapidly compress the
cold gas, enhancing its density while reducing its volume, until the cold
condensations acquire again the thermal pressure value of the hot gas. Given the
initially similar values of density ahead of and behind the shocks, one can show that
their velocity is only a function of the temperature $T$ of the hot gas
($V_\mathrm{RP} \approx [(kT)/(\mu m_p)]^{0.5}$, where $k$ is
the Boltzmann constant, $\mu m_p$ is the mean gas-particle mass, and $m_p$
is the proton mass), and thus cold parcels of gas within the SSC are rapidly driven
into small high density condensations.   

The continuous occurrence of thermal instabilities  results in the accumulation
of mass in this region and in a very chaotic, highly non-stationary
hydrodynamical pattern, with a number of radiative shocks and cooling fronts
propagating within  the cluster volume. The continuous accumulation of
thermally unstable matter must finally lead to its re-processing into further
generations of stars \citep{2005ApJ...628L..13T}. Unfortunately,  1D
simulations do not allow one to reach an adequate understanding of the physics
within massive clusters in the bimodal regime, nor to make realistic
predictions regarding the fate of the matter reinserted by massive stars.  In
order to develop a more realistic model, here we present detailed 2D
hydrodynamic simulations of the gaseous flows that result from the
thermalization of the kinetic energy supplied by stellar winds and supernovae
ejecta inside the volume occupied by the stellar cluster. We focus on clusters
evolving in the bimodal regime. The major result from these simulations is that
the central zones of clusters evolving in the bimodal regime accumulate large
amounts of matter in the form of warm ($T \sim 10^4$ K) high density
condensations  embedded into a hot plasma of much lower density. The amount of
accumulated matter  depends on the excess star cluster mechanical luminosity
over the threshold value and grows with time, unless the accumulation of
reinserted matter inside the stagnation radius is compensated by  secondary
star formation. Our results also show that the stagnation surface itself has a
very complicated dynamic morphology that continuously changes with time.
Nevertheless, the average radius of the stagnation surface remains close
to that predicted by 1D and semi-analytic calculations.   

The paper is organized as follows. Section 2  describes the numerical model
and discusses the input physics. Section 3 present the results from our
numerical simulations and compares them with the semi-analytical results and
previous 1D simulations. In section 4 we discuss our results and 
section 5 gives a  summary of our  findings.

\clearpage

\clearpage

\section{The numerical approach}

The numerical models presented here are based on the finite difference Eulerian
hydrodynamic code ZEUD3D v3.4.2 \citep{1992ApJS...80..753S}. All
simulations have been carried out on a spherical ($r$, $\theta$) grid,
symmetric along the $\phi$-coordinate. We have set the radial size of grid cells
$\Delta r$ proportional to the radial coordinate $r$ which ensures that  all
grid-cells have $\Delta r \sim r\Delta\theta$. Another advantage
of this radially scaled grid is that the resolution is higher at smaller
radii (inside the cluster) where thermal instabilities are expected.

The cooling routine accounts for extremely fast cooling both in the wind or
within the SSC volume. The change of internal energy $e$ due to cooling is
\begin{equation}
\label{coolrate}
\left.\frac{de}{dt}\right|_\mathrm{cool} = - n_i n_e \Lambda(T,Z) \ ,
\end{equation}
where $n_i$ and $n_e$ are ion and electron densities, respectively. We compute
them from the gas density $\rho$ as $n_i = n_e = \rho / (\mu_\mathrm{ion}
m_p)$, where $\mu_\mathrm{ion} m_p$ is the average ion mass.
We assume $\mu = 0.609$ and $\mu_\mathrm{ion} = 1.27$ for all computed models.
$\Lambda(T, Z)$ is the cooling function. We use Raymond \& Cox cooling function
which has been supplemented with new elements and tabulated by
\citet{1995MNRAS.275..143P}.

The RHS of equation~(\ref{coolrate}) is evaluated in the middle of time-steps to
maintain the second order accuracy of the code and the energy conservation
equation is then solved iteratively using the Brendt algorithm which is more
stable and accurate than the Newton-Raphson method originally used in ZEUS.

The cooling rate has to be included in the computation of the time-step, 
otherwise rapid cooling not resolved in time may lead to the occurrence of
negative temperatures. A common way of solving this is to limit the amount of 
internal energy which can be radiated away during one time-step by setting
\begin{equation}
dt_\mathrm{cool} = 
\epsilon \left| \frac{e}{\left.\frac{de}{dt}\right|_\mathrm{cool}} \right|
\end{equation}
where $\epsilon$ is a safety factor smaller than unity 
(see e.g.~\citealt{1997A&A...318..595S}; where $\epsilon = 0.3$ was used).
In this work we use $\epsilon = 0.25$.

The global time-step, $dt$, is computed as follows:
\begin{equation}
dt = \left\{
\begin{array}{lcl}
dt = dt_\mathrm{HD}, & \mathrm{for} & dt_\mathrm{cool} \ge dt_\mathrm{HD}; \\
dt = dt_\mathrm{cool},
& \mathrm{for} & dt_\mathrm{HD} > dt_\mathrm{cool} \ge \delta\times
dt_\mathrm{HD};\\
dt = \delta\times dt_\mathrm{HD}, & \mathrm{for} & \delta\times dt_\mathrm{HD} >
dt_\mathrm{cool}\ (\mathrm{local\ substeps}\ dt_\mathrm{sub} = dt_\mathrm{cool});
\end{array}
\right.
\end{equation}
where $dt_\mathrm{HD}$ is the "hydrodynamic" time-step resulting from the
Courant-Friedrich-Levi criterion and $\delta$ is the minimum fraction of
$dt_\mathrm{HD}$ to which the global time-step $dt$ is allowed to drop. If, due
to the cooling condition, a certain cell requires an even smaller time-step (\ie
$dt_\mathrm{cool} < \delta dt_\mathrm{HD}$), the energy equation is integrated
numerically in that cell, using $dt_\mathrm{sub} = dt_\mathrm{cool}$ but
assuming that during this time the density and temperature in the cell are not
substantially affected by interactions with neighboring cells. This ensures that
CPU time is not wasted in cells where a high time resolution is not required. To
determine a reasonable value of $\delta$ we ran several tests on a low
resolution grid ($150\times 56$) and found that there are no appreciable
differences for $\delta \lesssim 0.3$. Therefore, we use $\delta = 0.1$.

In order to simulate the effect of the stellar UV radiation field, in some
of the simulations (see Table~\ref{modtab}), we do not allow the gas
temperature to drop below $T_\mathrm{lim} = 10^4$~K. This  is equivalent to
assuming that there are sufficient UV photons to ionize the dense thermally
unstable matter which otherwise would cool to much lower temperatures. 

The wind source was modelled by a continuous replenishment of mass and internal
energy in all cells within the cluster volume, at rates $q_m =
(3\dot{M}_\mathrm{SC})/(4\pi R_\mathrm{SC}^3)$ and $q_e =
(3\dot{L}_\mathrm{SC})/(4\pi R_\mathrm{SC}^3)$, respectively.  The procedure
applied to each cell within the cluster volume at every time-step is: 
\begin{enumerate}
\item the density and the total energy in a given cell are saved to $\rho_\mathrm{old}$ and
$e_\mathrm{tot,old}$
\item the new mass is inserted $\rho_\mathrm{new} =
\rho_\mathrm{old} + (1 +A_\mathrm{noise} \zeta ) q_\mathrm{m} dt$, the velocity is corrected so that the
momentum is conserved $\mathbf{v}_\mathrm{new} =
\mathbf{v}_\mathrm{old}\rho_\mathrm{old}/\rho_\mathrm{new}$
\item the internal energy is corrected to conserve the total energy 
$e_\mathrm{i,mid} = e_\mathrm{tot,old} - \rho_\mathrm{new}\mathbf{v}^2_\mathrm{new}/2$
\item the new energy is inserted in
a form of internal energy $e_\mathrm{i,new} = e_\mathrm{i,mid} + (1 + A_\mathrm{noise} \zeta ) q_\mathrm{e}dt$
\end{enumerate}
where $\zeta$ is a pseudo-random number from the interval $(-1, 1)$ generated
each time it is used, and $A_\mathrm{noise}$ is the relative amplitude of the
noise. The inclusion of a noise term is necessary to break the artificial
spherical symmetry imposed by the initial conditions (see below). The model is
very robust with respect to $A_\mathrm{noise}$.  Test runs with
$A_\mathrm{noise} = 0.01, 0.05, 0.1$ and $0.5$ lead to very similar general
properties (mass fluxes at boundaries, number of fragments formed and their
approximate sizes) in all  models. The only noticeable difference is the
duration of the initial relaxation period required to break the initial
symmetry. We use $A_\mathrm{noise} = 0.1$ for  all models described in this
paper.

The boundary conditions are set open at both r-boundaries and periodic at both
$\theta$-boundaries. The open inner r-boundary allows the dense clumps that
are not ejected from the cluster (through its boundary at $R_\mathrm{SC}$) to
leave the computational domain. Otherwise, they would accumulate within the
cluster and eventually fill its whole volume; this happens for some models with
high $L_\mathrm{SC}$ and low $\eta$, even with the open inner r-boundary (see
Section~\ref{S_other}). It is unphysical, because the accumulated mass exceeds
the amount which can be ionized by the available UV photons; therefore it should
cool down, become gravitationally unstable and collapse into stars. Moreover,
the accumulated mass ultimately becomes so high that it would be gravitationally
unstable even at $T = 10^4$~K. However, determining the fate of the dense mass
properly would need a model of star formation which takes into account radiation
transfer and self-gravity: this is not included in our current numerical model.

Two different initial conditions are used for the two series of models presented
here (see Table~\ref{modtab}). Models 1 -- 6 and 12  with $R_\mathrm{SC}=10$~pc
have as initial condition  the spherically symmetric semi-analytical solution
with $L_\mathrm{SC} = L_\mathrm{crit}$ \citep{2004ApJ...610..226S}. Models 7 -
11 and 13, with $R_\mathrm{SC} = 3$~pc, use as initial condition the
semi-analytical solution with the appropriate $L_\mathrm{SC}$. As this is only
fully defined for $r > R_\mathrm{st}$, at $t = 0$ the central region $r <
R_\mathrm{st}$ is filled with a zero velocity, constant density and temperature
gas, with values equal to those at the stagnation radius
$\rho=\rho(R_\mathrm{st})$ and $T = T(R_\mathrm{st})$, respectively. The
advantage of this approach is that such conditions are closer to the
quasi-stationary state and therefore it takes a shorter time to reach it.

\section{Results}

We have calculated two series of models (see Table~\ref{modtab}). Models in the
first series all have a cluster radius $R_\mathrm{SC} = 10$~pc and an assumed
heating efficiency $\eta = 1$. The cluster mechanical luminosities considered 
are: $2.5\times 10^{41}$, $10^{42}$, $10^{43}$ and $10^{44}$~erg s$^{-1}$, which
result in values of $L_\mathrm{SC}/L_\mathrm{crit} = 0.5$, $2$, $20$ and
$200$, respectively. The model with $L_\mathrm{SC}/L_\mathrm{crit} = 2$ was
computed with three different numerical resolutions $150\times 56$, $300\times 112$
and $600\times 224$ to check and establish convergence. The computational  domain
extents radially from $R_\mathrm{IB} = 2$~pc (the inner boundary) to
$R_\mathrm{OB} = 30$~pc (the outer boundary) and from $\theta_\mathrm{LB} =
\pi/2 - 0.5$ (left boundary) to $\theta_\mathrm{RB} = \pi/2 +
0.5$ (right boundary) in the axial direction.

For the second series of models, we assume a more compact cluster and more
realistic cluster parameters by setting $R_\mathrm{SC} = 3$~pc and $\eta = 0.1$
or $0.3$. We ran 6 models with 3 different ratios $L_\mathrm{SC}/L_\mathrm{crit}
= 2.5$, $25$ and $250$. In these cases the radial extent of the grid goes
from $R_\mathrm{IB} = 0.5$~pc to $R_\mathrm{OB} = 10$~pc, and the axial extent
goes from $\pi/3$ to $2\pi/3$.

In all cases we assumed an adiabatic terminal velocity is $v_\mathrm{A,\infty}
\equiv \sqrt{\frac{2L_\mathrm{SC}}{\dot{M}_\mathrm{SC}}}= 1000$~km/s, and the
lower  temperature limit was set equal to  $T_\mathrm{lim} = 10^4$~K, or 10$^2$
K (models~12 and 13). In cases with an $\eta < 1$, $v_\mathrm{A,\infty}$, was replaced by
$v_\mathrm{A,\infty} \equiv \sqrt{\frac{2 \eta
L_\mathrm{SC}}{\dot{M}_\mathrm{SC}}}$. Note that in all cases, radiative
cooling lowers the outflow velocity at the grid outer boundary to  somewhat
smaller values. 

\clearpage
\begin{table}
\begin{tabular}{|l|l|l|l|l|l|l|l|l|l|l|}
\hline
No. & 
$\frac{L_\mathrm{SC}}{\mathrm{[erg/s]}}$  & 
$\frac{R_\mathrm{SC}}{\mathrm{[pc]}}$ &
$\eta$ & $\frac{L_\mathrm{SC}}{L_\mathrm{crit}}$ & 
grid &
$T_\mathrm{lim}$ &
$\frac{R_\mathrm{st}}{\mathrm{[pc]}}$ &
$\frac{T_\mathrm{st}}{\mathrm{[}10^6\mathrm{K]}}$ &
$\frac{\dot{M}_\mathrm{out}}{\dot{M}_\mathrm{SC}}$ &
$T_\mathrm{min}$ \\
\hline
1   & $2.5\times 10^{41}$ & 10  & 1 & 0.5 & $300\times 112$ & $10^4$ & 0 & $10.4$ & 1.00 & $10^4$ \\ 
2   & $10^{42}$ & 10 & 1   & 2    & $150\times 56 $ & $10^4$ & 6.056 & $10.4$ & 0.81 & $10^4$ \\ 
3   & $10^{42}$ & 10 & 1   & 2    & $300\times 112$ & $10^4$ & 6.056 & $10.4$ & 0.84 & $10^4$ \\ 
4   & $10^{42}$ & 10 & 1   & 2    & $600\times 224$ & $10^4$ & 6.056 & $10.4$ & 0.79 & $10^4$ \\ 
5   & $10^{43}$ & 10 & 1   & 20   & $300\times 112$ & $10^4$ & 8.991 & $10.4$ & 0.39 & $10^4$ \\ 
6   & $10^{44}$ & 10 & 1   & 200  & $300\times 112$ & $10^4$ & 9.696 & $10.4$ & 0.19 & $10^4$ \\ 
\hline                                                      
7   & $10^{39}$ & 3  & 0.1 & 2.5  & $300\times 104$ & $10^4$ & 1.872 & $1.10$ & 0.77& $10^4$ \\ 
8   & $10^{40}$ & 3  & 0.3 & 2.5  & $300\times 104$ & $10^4$ & 1.860 & $3.48$ & 0.84& $10^4$ \\ 
9   & $10^{40}$ & 3  & 0.1 & 25   & $300\times 104$ & $10^4$ & 2.709 & $1.10$ & 0.98& $10^4$ \\ 
10  & $10^{41}$ & 3  & 0.3 & 25   & $300\times 104$ & $10^4$ & 2.707 & $3.48$ & 0.83& $10^4$ \\ 
11  & $10^{41}$ & 3  & 0.1 & 250  & $300\times 104$ & $10^4$ & 2.912 & $1.10$ & 1.00& $10^4$ \\ 
\hline
12  & $10^{43}$ & 10 & 1   & 20   & $300\times 112$ & $10^2$ & 8.991 & $10.4$ & 0.38& $10^2$ \\ 
13  & $10^{40}$ & 3 & 0.1  & 25   & $300\times 104$ & $10^2$ & 2.709 & $1.10$ & 0.44& $10^2$ \\ 
\hline
\end{tabular}
\caption{The set of computed models. The values of $R_\mathrm{st}$ and
$T_\mathrm{st}$ were determined by means  of semi-analytic models, the values of
$\dot{M}_\mathrm{out}/\dot{M}_\mathrm{SC}$ are the mass
flux through the cluster border in the 2D simulations, averaged over the time
interval $0.1 - 0.8$~Myr for $R_\mathrm{SC} = 10$~pc models and $1-2$~Myr for
$R_\mathrm{SC} = 3$~pc models.}
\label{modtab}
\end{table}
\clearpage

\subsection{Model 1,  $L_\mathrm{SC}/L_\mathrm{crit} = 0.5$}

This model was computed in order to test the numerical code against the
semi-analytical solution which is known for $L_\mathrm{SC} < L_\mathrm{crit}$.
Despite the perturbations of the deposited mass and energy, the flow is
perfectly stationary. The resultant radial density, temperature and velocity
profiles are shown in Figure~\ref{m1d} where they are compared to the
semi-analytic solution. The agreement is very good, despite the fact that the
2-D model does not calculate the  central sphere of radius $R_\mathrm{IB}$, and
this induces a small discrepancy in the velocity in the inner cluster regions.

\subsection{Model 5, $L_\mathrm{SC}/L_\mathrm{crit} = 20$}

\begin{figure}
\plotone{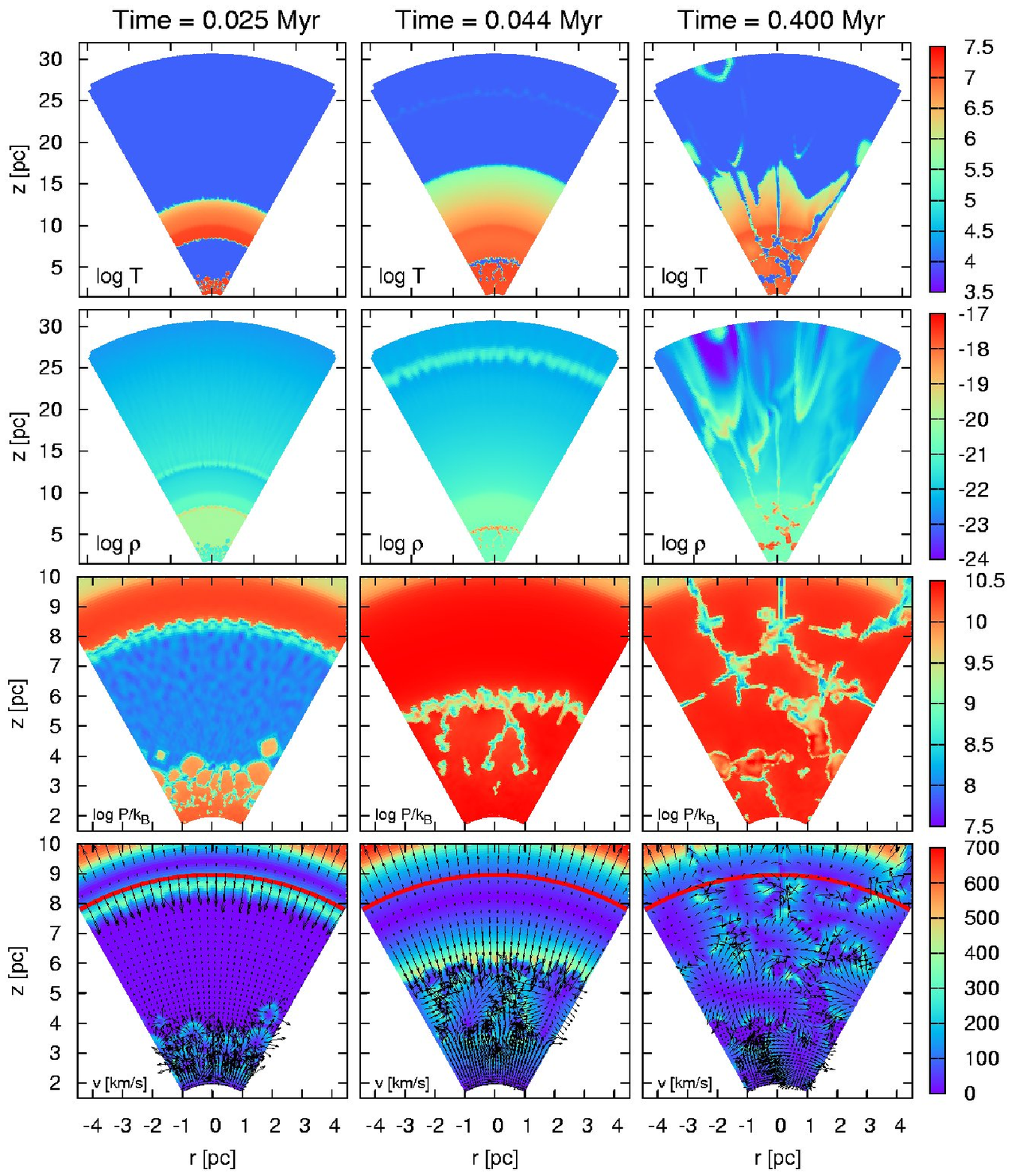}
\caption{Model~5 ($L_\mathrm{SC}/L_\mathrm{crit} = 20$, $\eta = 1$) at $t =
0.025$~Myr (left column), $t = 0.044$~Myr (middle column) and $t = 0.4$~Myr
(right column). The first two rows of panels show the logarithm of the wind
temperature and density, respectively, across the whole computational domain.
The third row shows the logarithm of pressure in the cluster central region
(below $R_\mathrm{SC} = 10$~pc), and the bottom row shows the wind velocity in
the same region, as arrows together with its magnitude coded by the color scale.
The red arc is the stagnation radius given by the semi-analytical solution.}
\label{rTd43}
\end{figure}

We continue with a detailed description of model~5 which exhibits the typical
hydrodynamic behavior for clusters above the threshold line, $L_\mathrm{SC} >
L_\mathrm{crit}$. The model starts with the semi-analytical solution for
$L_\mathrm{SC} = L_\mathrm{crit}$ and from then onwards, at each time step,  the
mass and energy input rates are consistent with the selected ratio
$L_\mathrm{SC}/L_\mathrm{crit} = 20$. Initially, the density  grows slowly all
over the cluster volume, and this steadily enhances the radiative cooling,
causing lower temperatures. Eventually, as  the temperature approaches $T\sim
3\times10^5$~K, the cooling rate increases steeply, and thermal instability
occurs. This  lowers the temperature rapidly to $T = 10^4$ K, particularly
in the densest central regions. On the other hand, the outer regions, where the
density is lower, remain hot  and thus  a large pressure gradient between  the
two regions leads to the formation of a strong shock wave propagating inwards
(see Figure~\ref{rTd43}, left column).

The simulations show a very dynamic evolution in which regions of hot gas
of different dimensions appear and grow close to the center, as more  energy is
deposited within the cluster volume.  The hot gas expands super-sonically into
the low pressure warm ($10^4$~K)  surroundings. This decreases locally the
density of the hot gas, {preventing its becoming} thermally unstable, while at
the same time the parcels of warm gas are compressed from all sides into high
density condensations, until they reach pressure equilibrium with the
surrounding hot gas. The hot regions grow until they again occupy most of
the cluster volume, see Figure~\ref{rTd43}, middle column.

While all this is happening, the inward propagating shock wave, overruns the
central region, accelerating inwards most of the high density condensations, and
these  eventually leave the computational grid as they cross the inner boundary.
As the high temperatures are  reestablished, the overall pressure gradient
vanishes and the shock wave decays. Once the hot gas again permeates the
cluster volume, the density starts to grow again and the whole process
repeats itself, but in a weaker form because of the presence of a few dense
condensations surviving from the previous cycle. This model exhibits 2-3
subsequent weaker periods of oscillations similar to the ones observed in 1D
simulations (\citealt{2007ApJ...658.1196T}; the low-energy model).

Finally, the oscillations vanish and a quasi-equilibrium situation is
established, in which high density condensations form at an approximately
constant rate (see Figure~\ref{rTd43}, right column). The amount of mass which goes into
warm dense condensations is just enough to maintain the hot medium close to the
thermally unstable regime. It is a quasi-stationary  situation: the density
tends to grow everywhere, surpassing some threshold value that favours the
thermal instability in the densest central regions. This sudden loss of pressure
prevents matter within the stagnation  boundary from escaping the
cluster as a wind. The thermally unstable gas is instead packed into high
density condensations, most of which leave the computational grid through the
inner boundary. The location of the stagnation surface, which separates the
region where the thermal instability occurs from the outer stationary  wind,
also experiences some oscillations. This leads to a non-spherical surface
with an average radius close to that  given by the analytic approximation
\citep{2007A&A...471..579W},
\begin{equation}
\frac{R_\mathrm{st}^3}{R_\mathrm{SC}^3} = 1 -
\left(\frac{L_\mathrm{crit}}{L_\mathrm{SC}}\right)^{1/2} \ .
\end{equation}
The $R_\mathrm{st}$ radius at which vicinity the orientation of the
velocity vectors abruptly changes from an outward to an inward motion has been
indicated with a red line in the bottom panels of Figure~\ref{rTd43}.

\subsection{Models 2-4, $L_\mathrm{SC}/L_\mathrm{crit} = 2$}

\begin{figure}
\plotone{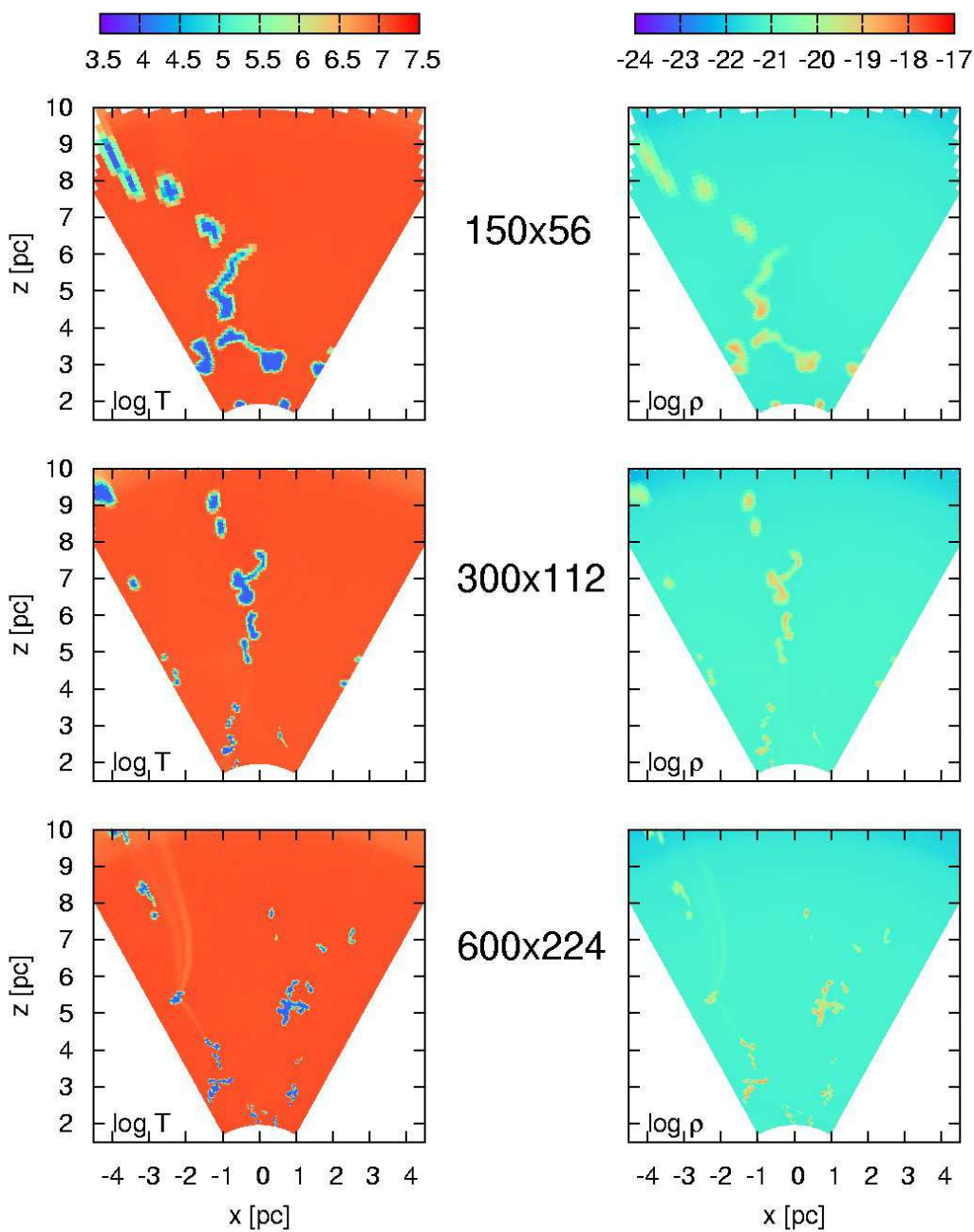}
\caption{Models~2-4 ($L_\mathrm{SC}/L_\mathrm{crit} = 2$, $\eta = 1$) at $t = 0.25$~Myr. 
The left and right columns show the logarithm of the wind temperature and
density, respectively, in the cluster central region. The three rows of panels
represent the different grid resolutions: the top row is $150\times56$
(model~2), the middle row $300\times112$ (model~3) and the bottom row
$600\times224$ (model~4). 
}
\label{rTd42}
\end{figure}

In these cases the stagnation radius $R_\mathrm{st}$ is smaller than in model~5,
resulting in a smaller thermally unstable region. The lower mass deposition rate
results in a smaller amount of high density gas being formed there (see
Figure~\ref{rTd42}). Otherwise, the evolution is similar to that of model~5,
including the initial relaxation period of intense thermal instability, the
appearance of re-pressurizing shocks which lead to high density condensations,
and the exit of most of these through the inner boundary, to finally reach
equilibrium between the formation of high density condensations and the mass
deposition rate.

We have computed this model on three different grids to check how the resolution
affects the results (see the different rows of Figure~\ref{rTd42}). Although the
resultant fragments tend to be smaller and more structured on the higher
resolution grids, the global characteristics such as the mass flux through the
cluster border (see Table~\ref{modtab}) are in a reasonable agreement.

\subsection{Model 6, $L_\mathrm{SC}/L_\mathrm{crit} = 200$}

\begin{figure}
\plotone{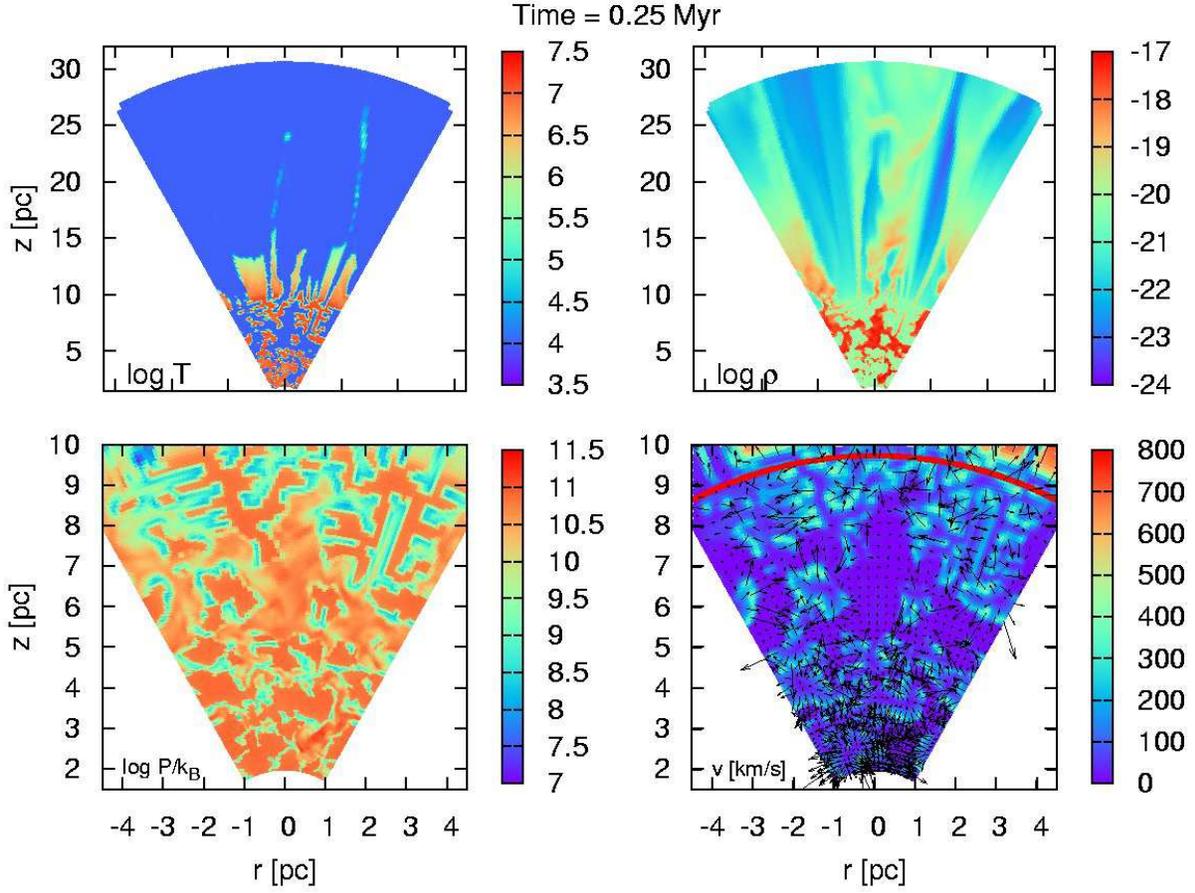}
\caption{Model~6 ($L_\mathrm{SC}/L_\mathrm{crit} = 200$, $\eta = 1$) 
at $t = 0.25$~Myr. The top left and right panels show the logarithm of the wind
temperature and density, respectively, across the whole computational domain.
The bottom panels show only the cluster region (below $R_\mathrm{SC} = 10$~pc),
the left panel represents the logarithm of pressure, the right panel shows 
the wind velocity as arrows together with its magnitude coded by the color 
scale. The red arc is the stagnation radius given by the semi-analytical 
solution.}
\label{rTd44}
\end{figure}

The hydrodynamical behavior of this case is very similar to that of model~5, the
only differences are quantitative: In this case many more high density
condensations  form and  occupy a larger fraction of the cluster volume (see
Figure~\ref{rTd44}). The quasi-stationary region, above the rapidly varying
stagnation surface, becomes at times very narrow and,  as it is repeatedly
perturbed by high density condensations leaving the cluster,  at times it is not
even a contiguous region.  Nevertheless, most of the thermally unstable gas
driven into high density condensations  stays within the cluster volume and
after some time, as in model~5,  goes through the inner boundary and disappears
from the computational domain.

\subsection{Other Models}
\label{S_other}

The most important parameter of our cluster model is the ratio
$L_\mathrm{SC}/L_\mathrm{crit}$, as this defines the location of the stagnation
surface and thus the relative sizes of the thermally  unstable region and of the
quasi-stationary outer wind region. We have performed many more simulations (see
Table 1), all presenting the same general features as in model~5. For example
models~7 and 8, ($L_\mathrm{SC}/L_\mathrm{crit} = 2.5$, $\eta = 0.1$ and $0.3$)
have  $L_\mathrm{SC}/L_\mathrm{crit}$ values close to those of models~2 - 4.
However, the smaller heating efficiency $\eta$ assumed in these cases, results
in a smaller temperature and pressure of the hot medium, and this leads to a
stationary wind that expands with a smaller velocity. The lower temperature of
the hot gas leads  to slower velocity re-pressurizing shocks. The lower ambient
pressure of the hot gas also leads to larger final sizes  of the high density
condensations and after some  time to their accumulation near the central zones
of the computational grid. This sooner or later prevents the exit of matter
through the inner boundary in case of models~9 and 10
($L_\mathrm{SC}/L_\mathrm{crit} = 25$, $\eta = 0.1$ and $0.3$) and model~11
($L_\mathrm{SC}/L_\mathrm{crit} = 250$, $\eta = 0.1$). We believe this is an
artifact promoted by the fact that we do not allow for this matter to go into
star formation.

\begin{figure}
\plotone{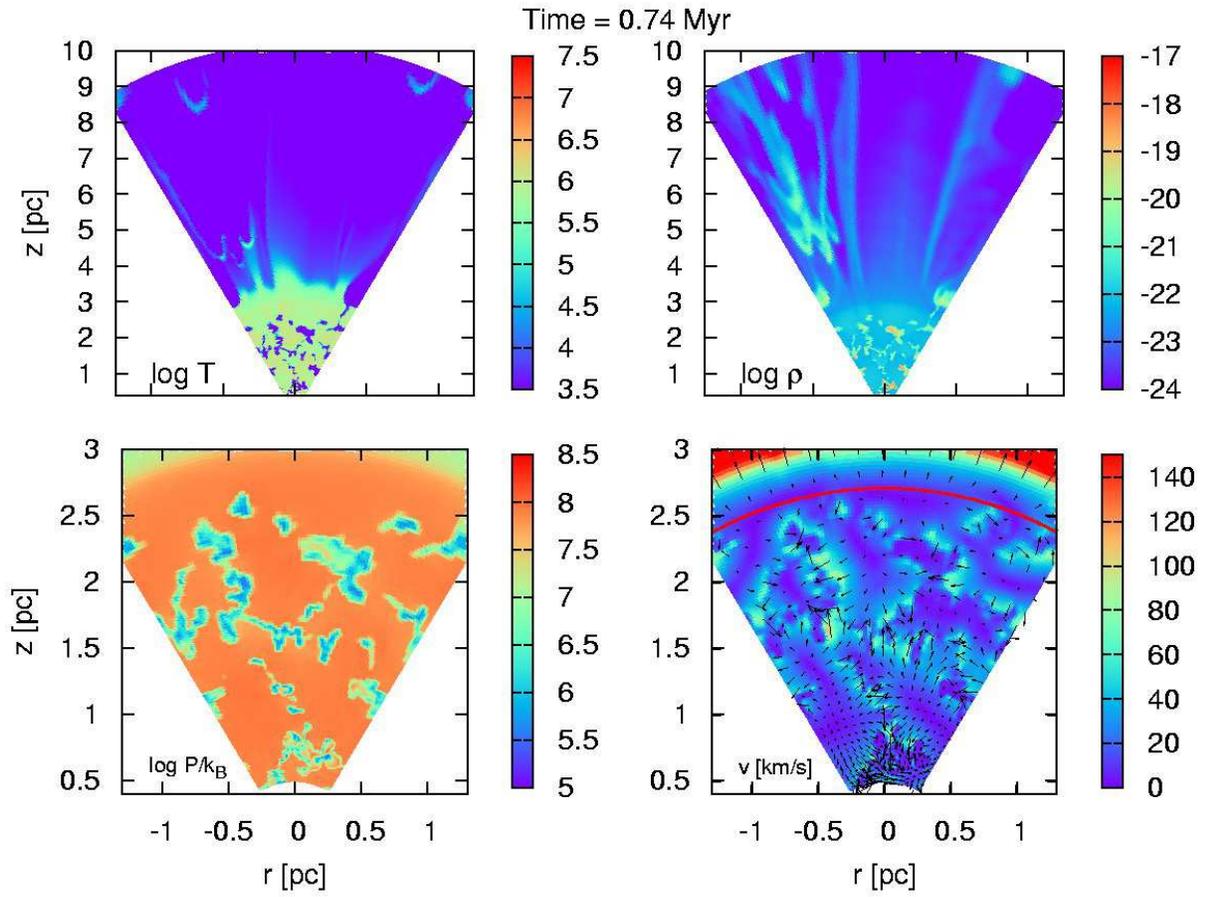}
\caption{Model~13 ($L_\mathrm{SC}/L_\mathrm{crit} = 25$, $\eta = 0.1$) 
at $t = 0.74$~Myr. The meaning of the panels is the same as in
Figure~\ref{rTd44}. 
}
\label{rTd_tm2}
\end{figure}

Models~12 and 13 are similar to models~5 and 9, respectively, but the gas
is allowed to cool to $10^2$~K instead of $10^4$~K. A comparison between
model~12 and 5 shows very similar results. This means that the reduced volume
of clumps in model~12 (due to their lower temperature and hence pressure) does
not affect the ratio of the clumps that are ejected from the cluster to the
clumps that stay there and finally leave the computational domain through the
inner boundary. Therefore, the outflow from the cluster ($\dot{M}_\mathrm{out}$)
remains the same in both models. However, in the case of model~13
($R_\mathrm{SC} = 3$~pc, $\eta = 0.1$), the reduction of the clump sizes
prevents accumulating and filling the cluster with a dense warm material, as
happens in model~9. As shown in Figure~\ref{rTd_tm2} and in Table~\ref{modtab}
the formation of new clumps via thermal instability is compensated with the
outflow from the computational zone making, model~13 more realistic.

\clearpage

\begin{figure}
\plotone{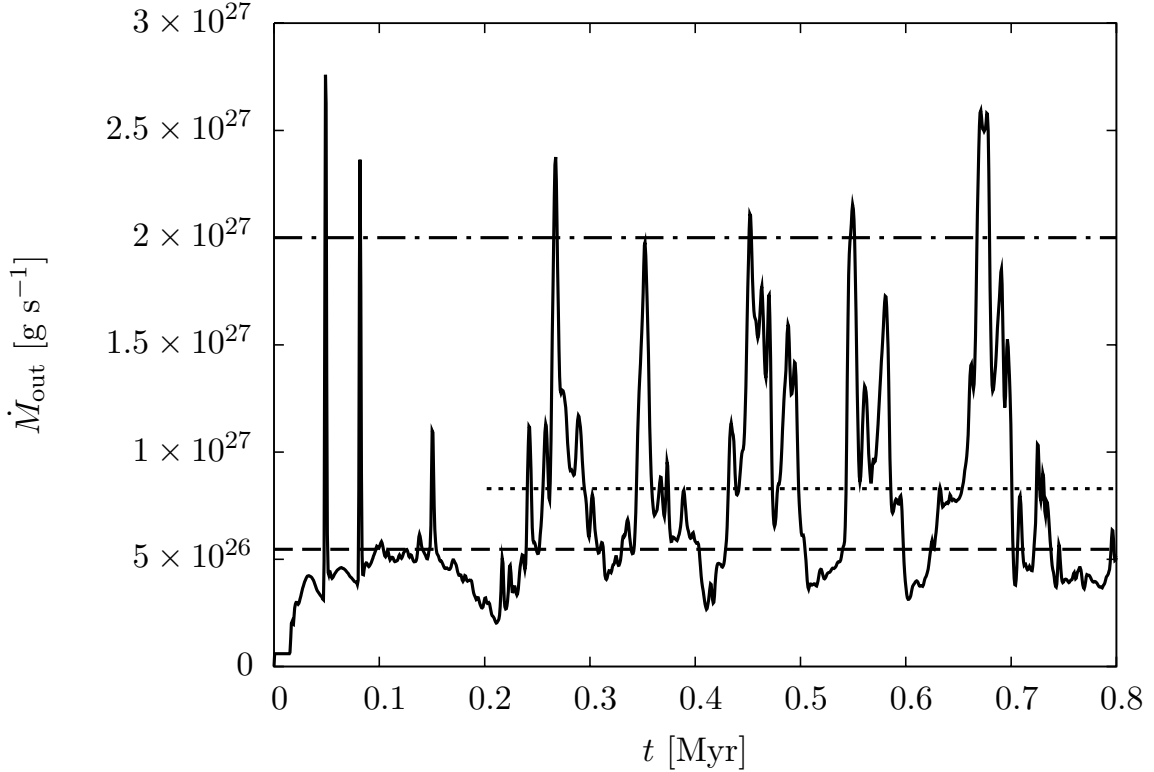}
\caption{
The solid line represents the time evolution of the mass flux measured at the
outer boundary of the computation domain in model~5. The mass deposition
rate to the whole cluster $\dot{M}_\mathrm{SC}$ and its fraction deposited
between $R_\mathrm{st}$ and $R_\mathrm{SC}$ are represented by dash-dotted line and
dashed line, respectively. The dotted line represents the average flux at the
outer boundary in the period 0.2 - 0.8 Myr. Those average mass fluxes are shown
as symbols in Figure~\ref{dMOB}. The rarefied wind produces the flux close to the
value given by the dashed line. The maxima of $\dot{M}_\mathrm{out}$ are due to condensations
passing through the outer boundary. They are preceded by short periods in
which $\dot{M}_\mathrm{out}$ drops below the dashed line, as if the condensations 
"cast shadows" to the regions above them -- the mass flux of the rarefied wind
is slightly lower then and corresponds to the moments when the outer
boundary is shadowed by an approaching condensation(s).
}
\label{fluxes43}
\end{figure}

\begin{figure}
\plotone{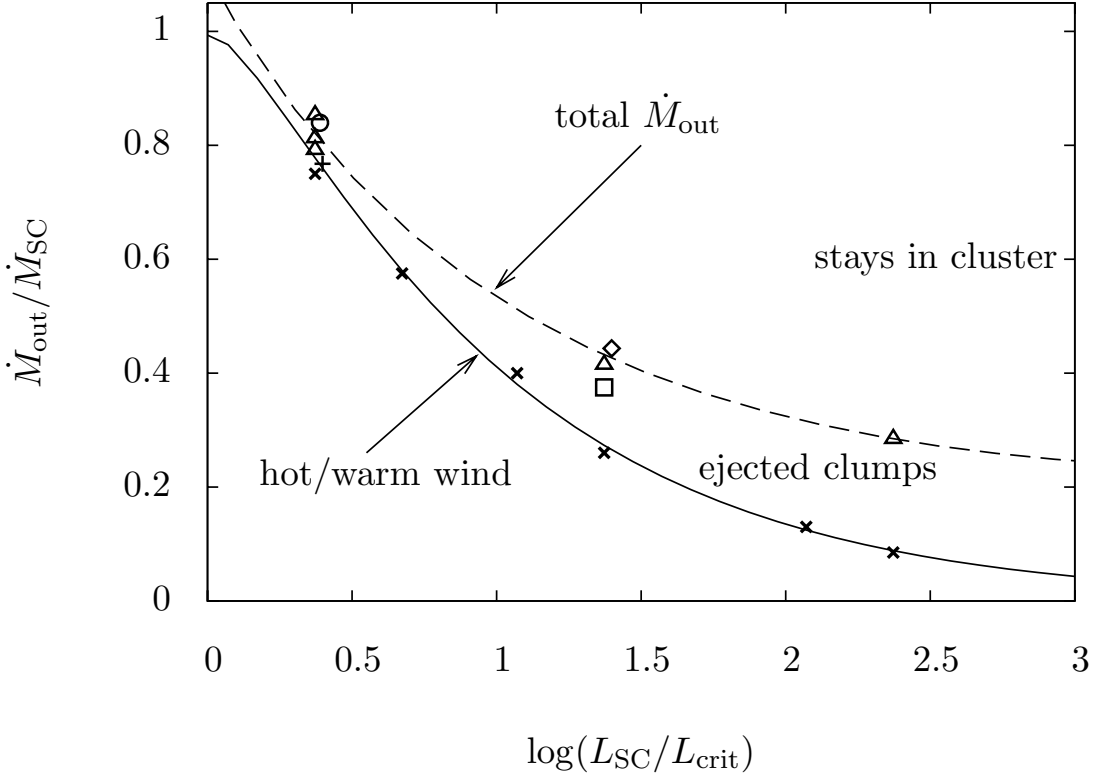}
\caption{The fraction of the matter inserted within the stellar cluster which is
able to leave it as a function of the cluster luminosity normalized by its
threshold value ($L_\mathrm{SC}/L_\mathrm{crit}$). The solid line represents the
fraction of mass which is inserted above $R_\mathrm{st}$ and which leaves the
cluster in a form of the hot wind (which may eventually cool down outside the
cluster). The $\times$ symbols denote the 1D simulations described in
\citet{2007ApJ...658.1196T}. The other symbols represent the 2D
simulations: triangles (models 2 -- 6), the plus (model 7),
the circle (model~8), the square (model~12), and the diamond (model~13). 
The values were obtained by averaging over time periods $0.2-0.8$~Myr (models 2
-- 6), and $1-2$~Myr (models~7, 8, 12 and 13). The dashed line is the fit to the
outflows of 2D models.} 
\label{dMOB}
\end{figure}
\clearpage

\section{Discussion}

Figure~\ref{fluxes43}  shows the time evolution of the mass flux through the
outer boundary of the computational domain for  model~5. The outflow becomes
quasi-stationary as an equilibrium is established between the formation of high density
condensations and their ejection from the cluster and exit through the inner
boundary. The average mass flux across the outer boundary of the
computation domain (dotted line in Figure~\ref{fluxes43}) slightly exceeds the rate of mass
deposition that occurs between the stagnation radius $R_\mathrm{st}$ and the star
cluster edge $R_\mathrm{SC}$. This is because some dense condensations  that
formed in the thermally unstable inner region ($r < R_\mathrm{st}$) cross the
stagnation surface and eventually leave the cluster contributing to the total
mass flux across the outer boundary of the computational grid.

Figure~\ref{dMOB} shows how the relative outflow from the cluster (measured as
the mass flux at the outer boundary, averaged over long time periods) depends on
the ratio $L_\mathrm{SC}/L_\mathrm{crit}$. We plot all the models where the
equilibrium between the clump formation and their removal through inner or outer
boundaries is reached. Other models (Nos 9, 10 and 11), where the volume of the
cluster is completely filled with the warm matter (see Sect. 3.5.) are omitted.
The outflow from the cluster consists of  two components: the hot wind, which
originates in the outer region of the cluster above $R_\mathrm{st}$, and the
warm dense clumps that are formed below $R_\mathrm{st}$ and that manage to pass
to the outer region where they are ejected from the cluster. The figure shows
three zones: the bottom zone is the fraction of the deposited mass
that goes into the wind, the middle zone represents the mass in ejected
high density condensations that stream away with the wind  ($\lesssim 20\%
\dot{M}_\mathrm{SC}$), and the upper zone is the mass which stays in the
cluster (available for star formation). Note that the hot wind also cools down
to temperatures $T \sim 10^4$ K at short distances from the cluster surface, so,
finally there are also two phases in the outflow: the warm wind and the dense
condensations (which should  expand unless they  cool even further).

In order to advance the problem further, apart from an adequate hydrodynamic
model that takes into consideration specific characteristics of SSCs (\ie their
high masses, small radii, large stellar densities and extreme output of
mechanical energy), one would need to couple the hydrodynamics to the UV
radiation field. We have assumed here that this is the case and that the  UV
radiation field generated by the massive stars evolving in the cluster keeps the
temperature of the thermally unstable gas at $T \sim 10^4$~K through
photoionization. This may be true for very young clusters, before the supernova
era starts ($t_\mathrm{SN} \sim $3 Myr). However, older clusters, with a reduced
ionizing photon flux, would soon  become unable to photoionize all the gas that
became thermally unstable. In such  cases, (see Figure~\ref{rTd_tm2})  the
thermally unstable gas would continue to cool further, while being compressed
into correspondingly smaller volumes.  As a result, the increasingly high
densities would trap the ionization front in the outer skins of the
condensations and their interior would remain neutral and at low temperature
($\sim 10$~K). In this way, if a parcel of gas with an original temperature
$\sim 10^7$ K, that becomes thermally unstable and is able to cool down to
$10$~K, would experience a rapid evolution in which its volume, in order to
preserve pressure equilibrium, would be reduced by six orders of magnitude while
its density would become six orders of magnitude larger.

Another factor not taken into account in the present set of simulations is
gravity. The gravitational pull caused by the cluster is perhaps not
significant for the high temperature gas, as this has a sound speed of several
hundreds of km$\,$s$^{-1}$, much larger that the escape speed from  the cluster.
However, it should promote a faster exit of low temperature condensations across
the inner grid  boundary. Indeed, if one considers a condensation that develops
at the stagnation radius, its free-fall time to the  cluster center will be
$\tau_\mathrm{ff} = \pi R_\mathrm{st}^{3/2} / (2GM_\mathrm{st})^{1/2}$ where
$M_\mathrm{st}$ is the mass below $R_\mathrm{st}$. In pc and solar mass units,
it is $\tau_\mathrm{ff} \approx 16.5 R_\mathrm{SC}^{3/2} /
M_\mathrm{SC}^{1/2}$~Myr which leads to time-scales much shorter than the
computational time. Thus gravity  would lead to an increase in the speed
of such condensations as they move to cross the inner boundary. The implication
of this is a faster accumulation of the thermally unstable matter near the center
of the cluster, where further generations of stars are to take place. Another
important factor, also promoting a faster matter accumulation and further
stellar generations arises from the self-gravity of the thermally unstable gas.

\section {Conclusions}

Here we have confirmed  by means of 2D hydrodynamic simulations the
existence of a bimodal solution for SSCs above the threshold line. We have shown
that the evolution within the volume defined by the stagnation surface is very
dynamic. The stagnation surface itself has a very  dynamic morphology that
continuously changes with time. Nevertheless, the average value of the stagnation
radius remains close to the value  predicted by  1D simulations and
semi-analytic solutions.   This region suffers a very dynamic evolution in which
parcels of gas continuously become thermally unstable and are rapidly driven
into very small volumes to compensate their sudden loss of pressure. The number
of such regions depends on the excess star cluster mechanical luminosity above
the threshold value. The fraction of the cluster volume occupied by the warm
medium depends on the balance between the formation of high density
condensations, via thermal instability, and their removal via secondary star
formation and/or their escape from the cluster. In our model, the secondary star
formation is partly accounted for with the exit of high density gas across  the
inner boundary. However, a better treatment in the future would be to implement
a more physical description of secondary star formation.

We have also shown that the growth of high density condensations within a SSC
volume, is strongly linked to  the parameter $\eta$, as this determines the
location  of the critical luminosity, $L_\mathrm{crit}$, in the star cluster
mechanical luminosity vs size plane, and thus determines how far above the
critical luminosity a cluster is. It also influences the sound speed (and
pressure) in the thermal instability region and  as a result it fixes the
final high density that condensations, confined by pressure, attain
(higher $\eta$ $\Rightarrow$ higher pressure $\Rightarrow$ denser
condensations).

We have shown that most of the condensations generated within the stagnation
volume, are unable to leave the cluster volume and thus accumulate.  Eventually
this will result in further generations of star formation. The central
implication of this result is that most of the metals processed by stars in
massive and compact SSCs will not be ejected back into the host-galaxy
ISM, an important issue to be taken into consideration by models of galactic
chemical evolution and $\Lambda$CDM models of the universe. Careful analysis of
observational data of the most massive and compact star clusters is now
required to select those which may evolve in the bimodal, catastrophic cooling
regime. For all of them we expect a mixture hot X-ray gas, a warm partially
photo-ionized plasma, and an ensemble of accumulating cool dense
condensations. Thus they should be detectable in the X-ray band, visible,
radio and infrared regimes simultaneously.

Note also, that the high stellar densities associated with SSCs resemble those
of globular clusters, in which  star-to-star abundance inhomogeneities have been
observed  \citep{2007ApJ...665.1164B}. This can be understood if globular
clusters have entered the bimodal regime during their early evolution,
and this has led to multiple stellar generations forming with the matter
reinserted into the cluster volume. Such clusters, having a reduced amount of
matter being lost back into the ISM, would have remained more stable against
disruption. 

Other open issues and some more astrophysical consequences of clusters
undergoing this bimodal evolution have been discussed by
\citet{2008Ap&SS.granada.jan}.

\acknowledgements

The authors would like to express their thanks to Anthony Whitworth for many
helpful discussions leading to a significant improvement of the paper. An
anonymous referee also provided valuable comments and suggestions. This study
has been supported by CONACYT - M\'exico, research grant 60333 and 47534-F and
AYA2004-08260-CO3-O1 from the Spanish Consejo Superior de Investigaciones
Cient\'\i{}ficas, by the Institutional Research Plan AV0Z10030501 of the Academy
of Sciences of the Czech Republic and by project LC06014 Center for Theoretical
Astrophysics of the Ministry of Education, Youth and Sports of the Czech
Republic. RW acknowledges support by the Human Resources and Mobility Programme
of the European Community under the contract MEIF-CT-2006-039802.

\bibliographystyle{aa}
\bibliography{winds2d}

\end{document}